# Blended RAG: Improving RAG (Retriever-Augmented Generation) Accuracy with Semantic Search and Hybrid Query-Based Retrievers

1st Kunal Sawarkar
*IBM*
Kunal@ibm.com

2nd Abhilasha Mangal
*IBM*
Abhilasha.Mangal@ibm.com

3rd Shivam Raj Solanki
*IBM*
Shivam.Raj.Solanki@ibm.com

*Abstract*—Retrieval-Augmented Generation (RAG) is a prevalent approach to infuse a private knowledge base of documents with Large Language Models (LLM) to build Generative Q&A (Question-Answering) systems. However, RAG accuracy becomes increasingly challenging as the corpus of documents scales up, with Retrievers playing an outsized role in the overall RAG accuracy by extracting the most relevant document from the corpus to provide context to the LLM. In this paper, we propose the 'Blended RAG' method of leveraging semantic search techniques, such as Dense Vector indexes and Sparse Encoder indexes, blended with hybrid query strategies. Our study achieves better retrieval results and sets new benchmarks for IR (Information Retrieval) datasets like NQ and TREC-COVID datasets. We further extend such a 'Blended Retriever' to the RAG system to demonstrate far superior results on Generative Q&A datasets like SQUAD, even surpassing fine-tuning performance.

## I. INTRODUCTION

RAG represents an approach to text generation that is based not only on patterns learned during training but also on dynamically retrieved external knowledge. This method combines the creative flair of generative models with the encyclopedic recall of a search engine. The efficacy of the RAG system relies fundamentally on two components: the Retriever (R) and the Generator (G), the latter representing the size and type of LLM.

The language model can easily craft sentences, but it might not always have all the facts. This is where the Retriever (R) steps in, quickly sifting through vast amounts of documents to find relevant information that can be used to inform and enrich the language model's output. Think of the retriever as a researcher part of the AI, which feeds the contextually grounded text to generate knowledgeable answers to Generator (G). Without the retriever, RAG would be like a well-spoken individual who delivers irrelevant information.

## II. RELATED WORK

Search has been a focal point of research in information retrieval, with numerous studies exploring various methodologies. Historically, the BM25 (Best Match) algorithm, which uses similarity search, has been a cornerstone in this field, as explored by Robertson and Zaragoza (2009). [1]. BM25 prioritizes documents according to their pertinence to a query, capitalizing on Term Frequency (TF), Inverse Document Frequency (IDF), and Document Length to compute a relevance score.

Dense vector models, particularly those employing KNN (k Nearest Neighbours) algorithms, have gained attention for their ability to capture deep semantic relationships in data. Studies by Johnson et al. (2019) demonstrated the efficacy of dense vector representations in large-scale search applications. The kinship between data entities (including the search query) is assessed by computing the vectorial proximity (via cosine similarity etc.). During search execution, the model discerns the 'k' vectors closest in resemblance to the query vector, hence returning the corresponding data entities as results. Their ability to transform text into vector space models, where semantic similarities can be quantitatively assessed, marks a significant advancement over traditional keyword-based approaches. [2]

On the other hand, sparse encoder based vector models have also been explored for their precision in representing document semantics. The work of Zaharia et al. (2010) illustrates the potential of these models in efficiently handling high-dimensional data while maintaining interpretability, a challenge often faced in dense vector representations. In Sparse Encoder indexes the indexed documents, and the user's search query maps into an extensive array of associated terms derived from a vast corpus of training data to encapsulate relationships and contextual use of concepts. The resultant expanded terms for documents and queries are encoded into sparse vectors, an efficient data representation format when handling an extensive vocabulary.

### A. Limitations in the current RAG system

Most current retrieval methodologies employed in Retrieval-Augmented Generation (RAG) pipelines rely on keyword and similarity-based searches, which can restrict the RAG system's overall accuracy. Table 1 provides a summary of the current benchmarks for retriever accuracy.

TABLE I: Current Retriever Benchmarks

| Dataset | Benchmark Metrics | NDCG@10 | p@20 | F1 |
|---|---|---|---|---|
| NQDataset | P@20 | 0.633 | 86 | 79.6 |
| Trec Covid | NDCG@10 | 80.4 | | |
| HotpotQA | F1 , EM | | | 0.85 |

While most of prior efforts in improving RAG accuracy is on G part, by tweaking LLM prompts, tuning etc.,[9] they have limited impact on the overall accuracy of the RAG system, since if R part is feeding irreverent context then answer would be inaccurate. Furthermore, most retrieval methodologies employed in RAG pipelines rely on keyword and similarity-based searches, which can restrict the system's overall accuracy.

Finding the best search method for RAG is still an emerging area of research. The goal of this study is to enhance retriever and RAG accuracy by incorporating Semantic Search-Based Retrievers and Hybrid Search Queries.

## III. BLENDED RETRIEVERS

For RAG systems, we explored three distinct search strategies: keyword-based similarity search, dense vector-based, and semantic-based sparse encoders, integrating these to formulate hybrid queries. Unlike conventional keyword matching, semantic search delves into the nuances of a user's query, deciphering context and intent. This study systematically evaluates an array of search techniques across three primary indices: BM25 [3] for keyword-based, KNN [4] for vector-based, and Elastic Learned Sparse Encoder (ELSER) for sparse encoder-based semantic search.

1) BM25 Index: The BM25 index is adept at employing full-text search capabilities enhanced by fuzzy matching techniques, laying the groundwork for more sophisticated query operations.
2) Dense Vector Index: We construct a dense vector index empowered by sentence transformers. It identifies the proximity of vector representations derived from document and query content.
3) Sparse Encoder Index: The Sparse EncodeR Retriever Model index is an amalgam of semantic understanding and similarity-based retrieval to encapsulate the nuanced relationships between terms, thereby capturing a more authentic representation of user intent and document relevance.

### A. Methodology

Our methodology unfolds in a sequence of progressive steps, commencing with the elementary match query within the BM25 index. We then escalate to hybrid queries that amalgamate diverse search techniques across multiple fields, leveraging the multi-match query within the Sparse Encoder-Based Index. This method proves invaluable when the exact location of the query text within the document corpus is indeterminate, hence ensuring a comprehensive match retrieval.

The multi-match queries are categorized as follows:
- Cross Fields: Targets concurrence across multiple fields
- Most Fields: Seeks text representation through different lenses across various fields.
- Best Fields: Pursues the aggregation of words within a singular field.
- Phrase Prefix: Operates similarly to Best Fields but prioritizes phrases over keywords.

After initial match queries, we incorporate dense vector (KNN) and sparse encoder indices, each with their bespoke hybrid queries. This strategic approach synthesizes the strengths of each index, channeling them towards the unified goal of refining retrieval accuracy within our RAG system. We calculate the top-k retrieval accuracy metric to distill the essence of each query type.

In Figure 1, we introduce a scheme designed to create Blended Retrievers by blending semantic search with hybrid queries.

### B. Constructing RAG System

From the plethora of possible permutations, a select sextet (top 6) of hybrid queries—those exhibiting paramount retrieval efficacy—were chosen for further scrutiny. These queries were then subjected to rigorous evaluation across the benchmark datasets to ascertain the precision of the retrieval component within RAG. The sextet queries represent the culmination of retriever experimentation, embodying the synthesis of our finest query strategies aligned with various index types. The six blended queries are then fed to generative question-answering systems. This process finds the best retrievers to feed to the Generator of RAG, given the exponential growth in the number of potential query combinations stemming from the integration with distinct index types.

The intricacies of constructing an effective RAG system are multi-fold, particularly when source datasets have diverse and complex landscapes. We undertook a comprehensive evaluation of a myriad of hybrid query formulations, scrutinizing their performance across benchmark datasets, including the Natural Questions (NQ), TREC-COVID, Stanford Question Answering Dataset (SqUAD), and HotPotQA.

## IV. EXPERIMENTATION FOR RETRIEVER EVALUATION

We used top-10 retrieval accuracy to narrow down the six best types of blended retrievers (index + hybrid query) for comparison for each benchmark dataset.

*1) Top-10 retrieval accuracy on the NQ dataset :* For the NQ dataset [5], our empirical analysis has demonstrated the superior performance of hybrid query strategies, attributable to the ability to utilize multiple data fields effectively. In Figure 2, our findings reveal that the hybrid query approach employing the **Sparse Encoder with Best Fields** attains the highest retrieval accuracy, reaching an impressive 88.77%. This result surpasses the efficacy of all other formulations, establishing a new benchmark for retrieval tasks within this dataset.

*2) Top-10 Retrieval Accuracy on TREC-Covid dataset:* For the TREC-COVID dataset [6], which encompasses relevancy scores spanning from -1 to 2, with -1 indicative of irrelevance and 2 denoting high relevance, our initial assessments targeted documents with a relevancy of 1, deemed partially relevant.

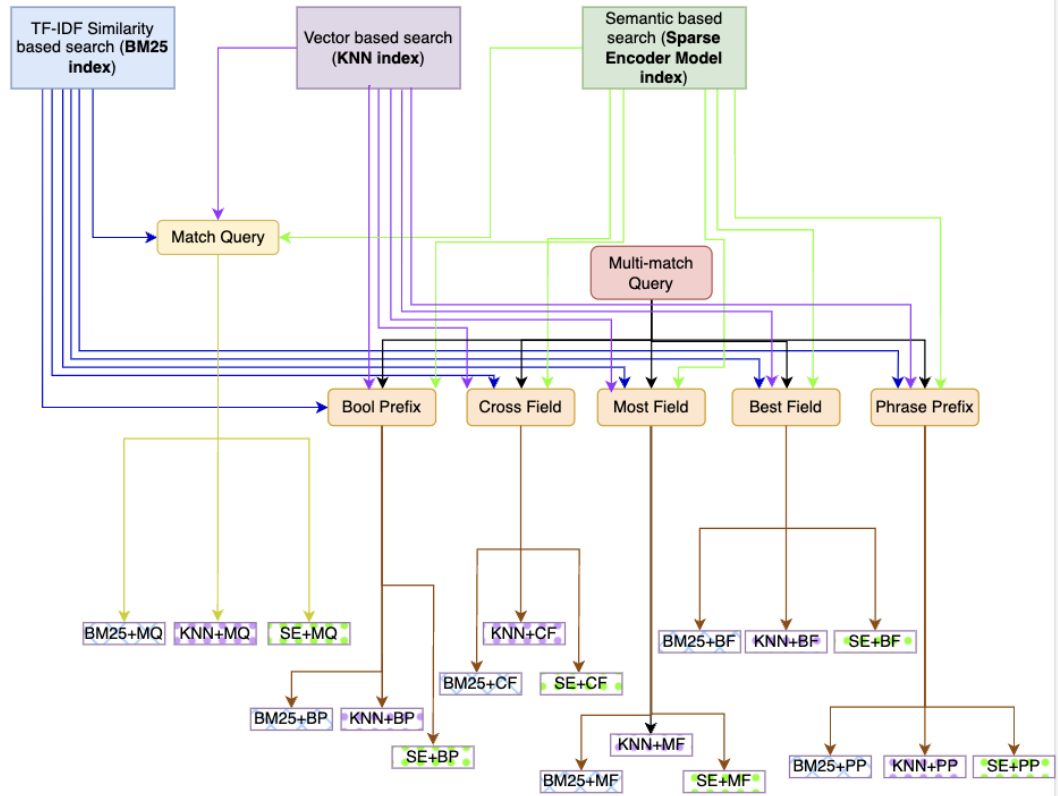

Fig. 1: Scheme of Creating Blended Retrievers using Semantic Search with Hybrid Queries.

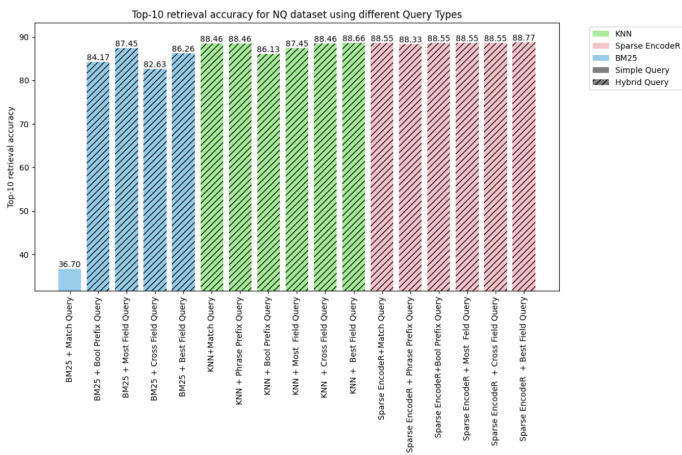

Fig. 2: Top-10 Retriever Accuracy for NQ Dataset

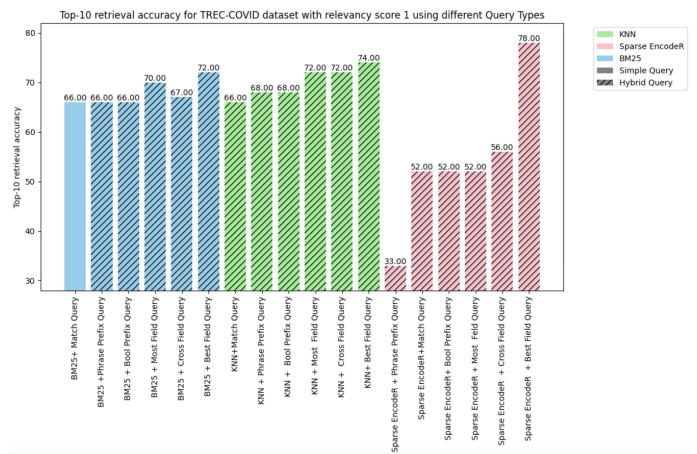

Fig. 3: Top 10 retriever accuracy for Trec-Covid Score-1

Figure 3 analysis reveals a superior performance of vector search hybrid queries over those based on keywords. In particular, hybrid queries that leverage the **Sparse EncodeR utilizing Best Fields** demonstrate the highest efficacy across all index types at 78% accuracy.

Subsequent to the initial evaluation, the same spectrum of queries was subjected to assessment against the TREC-COVID dataset with a relevancy score of 2, denoting that the documents were entirely pertinent to the associated queries. Figure 4 illustrated with a relevance score of two, where documents fully meet the relevance criteria for associated queries, reinforce the efficacy of vector search hybrid queries over conventional keyword-based methods. Notably, the hybrid query incorporating **Sparse Encoder with Best Fields** demonstrates a 98% top-10 retrieval accuracy, eclipsing all other formulations. This suggests that a methodological pivot towards more nuanced blended search, particularly those that effectively utilize the Best Fields, can significantly enhance retrieval outcomes in information retrieval (IR) systems.

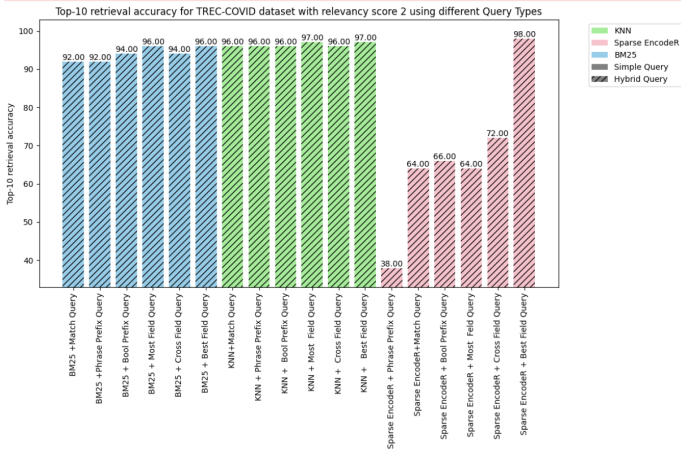

Fig. 4: Top 10 retriever accuracy for Trec-Covid Score-2

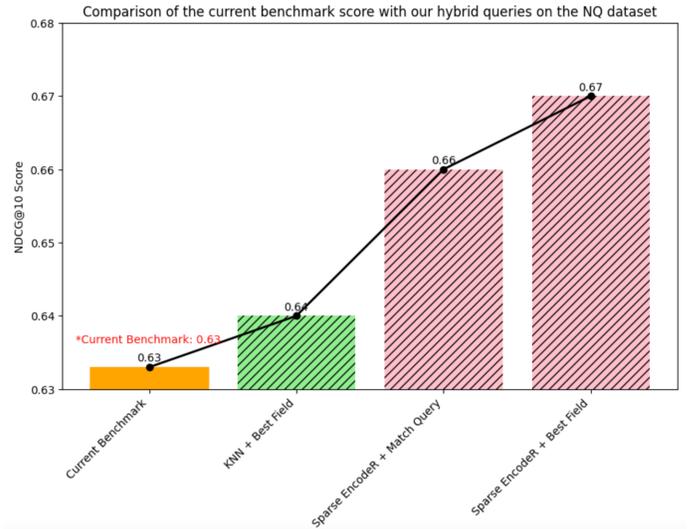

Fig. 6: NQ dataset Benchmarking using NDCG@10 Metric

TABLE II: Retriever Benchmarking using NDCG@10 Metric

| Dataset | Model/Pipeline | NDCG@10 |
| --- | --- | --- |
| Trec-covid | COCO-DR Large | 0.804 |
| Trec-covid | Blended RAG | 0.87 |
| NQ dataset | monoT5-3B | 0.633 |
| NQ dataset | Blended RAG | 0.67 |

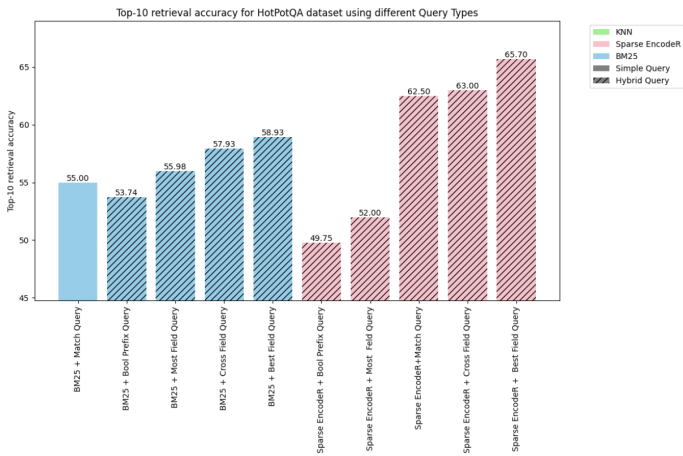

Fig. 5: Top 10 retriever accuracy for HotPotQA dataset

*3) Top-10 Retrieval Accuracy on the HotPotQA dataset:* The HotPotQA [7] dataset, with its extensive corpus of over 5M documents and a query set comprising 7,500 items, presents a formidable challenge for comprehensive evaluation due to compute requirements. Consequently, the assessment was confined to a select subset of hybrid queries. Despite these constraints, the analysis provided insightful data, as reflected in the accompanying visualization in Figure 5.

Figure 5 shows that hybrid queries, specifically those utilizing Cross Fields and Best Fields search strategies, demonstrate superior performance. Notably, the hybrid query that blends Sparse EncodeR with Best Fields queries achieved the highest efficiency, of 65.70% on the HotPotQA dataset.

### A. Retriever Benchmarking

Now that we have identified the best set of combinations of Index + Query types, we will use these sextet queries on IR datasets for benchmarking using NDCG@10 [8] scores (Normalised Discounted Cumulative Gain metric).

*1) NQ dataset benchmarking:* The results for NDCG@10 using sextet queries and the current benchmark on the NQ dataset are shown in the chart Figure 7. Our pipeline provides the best NDCG@10 score of 0.67, which is 5.8% higher than the current benchmark score of 0.633 achieved by the monoT5-3B model. Table II shows that all semantic search-based hybrid queries outperform the current benchmark score, which indicates that our hybrid queries are a better candidate for developing the RAG pipeline.

*2) TREC-Covid Dataset Benchmarking:* In our research, the suite of hybrid queries devised has demonstrably exceeded the current benchmark of 0.80 NDCG@10 score, signaling their superior candidature for the RAG pipeline. Figure 7 shows the results for NDCG@10 using sextet queries. Blended Retrievers achieved an NDCG@10 score of 0.87, which marks an 8.2% increment over the benchmark score of 0.804 established by the COCO-DR Large model (Table II).

*3) SqUAD Dataset Benchmarking:* The SqUAD (Stanford Question Answering Dataset) [9] is not an IR dataset, but we evaluated the retrieval accuracy of the SquAD dataset for consistency. Firstly, we created a corpus from the SqUAD dataset using the title and context fields in the dataset. Then, we indexed the corpus using BM25, dense vector, and Sparse Encoder. The top-k (k=5,10, and 20) retrieval accuracy results for the SqUAD dataset are calculated. Table III illustrates that for SQuAD, dense vector (KNN)-based semantic searches achieve higher accuracy than sparse vector-based semantic searches and traditional similarity-based searches, particularly for top-k retrieval performance with k values of 5, 10, and 20. (See Appendix for more details)

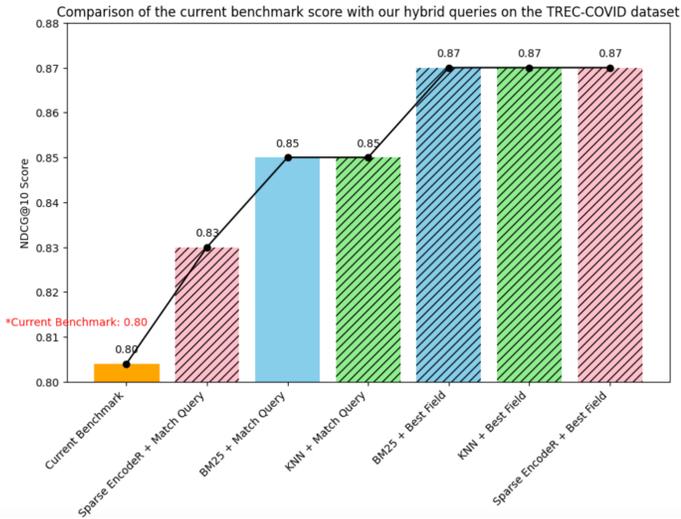

Fig. 7: TREC-Covid Dataset Benchmarking using NDCG@10 Metric

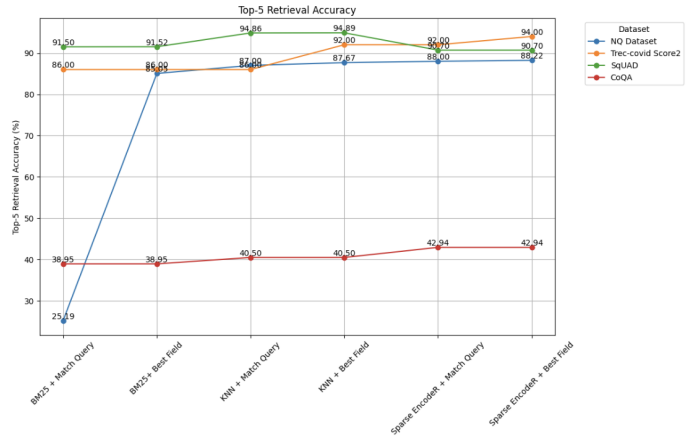

Fig. 8: Top-5 Retrieval Accuracy across Datasets

### B. Summary of Retriever Evaluation

We evaluated the retrieval accuracy using our approach, quantified by Top-k metrics where $k \in \{5, 10, 20\}$, across NQ, TREC-COVID, SQUAD, and CoQA datasets. This synopsis demonstrates the capability of our **Blended Retrieval** methodology within diverse informational contexts. Key observations are

- Enhanced retrieval accuracy is exhibited in all datasets except for CoQA [10]. This enhancement is attributable to the capability of our hybrid queries to effectively utilize available metadata to source the most pertinent results.
- Implementing dense vector-based (KNN) semantic search results in a marked improvement over keyword-based search approaches.
- Employing semantic search-based hybrid queries realizes better retrieval precision compared to all conventional keyword-based or vector-based searches.
- Furthermore, it is discernible that the Sparse Encoder-based semantic search, when amalgamated with the 'Best Fields' hybrid query, often provides superior results than any other method.

## V. RAG Experimentation

From the retriever evaluation experiments, we know the best retriever, i.e., the best combination of indices + query. In this section, we extend this knowledge to evaluate the RAG pipeline. To avoid the effect of LLM size or type, we perform all experiments using FLAN-T5-XXL.

### A. RAG Evaluation on the SqUAD Dataset

SqUAD is a commonly bench-marked dataset for RAG systems or Generative Q&A using LLMs. Our study juxtaposes three variations of the RAG pipeline from prior work using the evaluation metrics of Exact Match (EM) and F1 scores to gauge the accuracy of answer generation, as well as Top-5 and Top-10 for retrieval accuracy.

- RAG-original [11]: This variant, a model fine-tuned on the Natural Questions dataset, has been appraised without domain-specific adaptation.
- RAG-end2end [11]: As an extension of RAG-original, this model undergoes additional fine-tuning, tailored for domain adaptation to the SQuAD.
- Blended RAG: Distinctively, our Blended RAG variant has not undergone training on the SQuAD dataset or any related corpora. It harnesses an optimized amalgamation of field selections and hybrid query formulations with semantic indices to feed LLMs to render the most precise responses possible.

Consequently, as shown in Table IV, our Blended RAG showcases enhanced performance for Generative Q&A with F1 scores higher by 50%, even without dataset-specific fine-tuning. This characteristic is particularly advantageous for large enterprise datasets, where fine-tuning may be impractical or unfeasible, underscoring this research's principal application.

### B. RAG Evaluation on the NQ Dataset

Natual Questions (NQ) is another commonly studied dataset for RAG. The Blended RAG pipeline, utilizing zero-shot learning, was evaluated to ascertain its efficacy against other non-fine-tuned models. The assessment focused on the following metrics: Exact Match (EM), F1 Score, and retrieval accuracy (Top-5 and Top-20) in Table V.

Blended RAG (Zero-shot): Demonstrated superior performance with an EM of 42.63, improving the prior benchmark by 35%.

TABLE III: Blended Retriever Performance SqUAD Dataset

| SqUAD | BM25+MQ | BM25+BF | KNN+MQ | KNN+BF | SPARSE ENCODER+MQ | SPARSE ENCODER+BF |
|---|---|---|---|---|---|---|
| Top-5 | 91.5 | 91.52 | 94.86 | 94.89 | 90.7 | 90.7 |
| Top-10 | 94.43 | 94.49 | 97.43 | 97.43 | 94.13 | 94.16 |
| Top-20 | 96.3 | 96.36 | 98.57 | 98.58 | 96.49 | 96.52 |

TABLE IV: Evaluation of the RAG Pipeline on the SquAD Dataset

| Model/Pipeline | EM | F1 | Top-5 | Top-20 |
|---|---|---|---|---|
| RAG-original | 28.12 | 39.42 | 59.64 | 72.38 |
| RAG-end2end | 40.02 | 52.63 | 75.79 | 85.57 |
| Blended RAG | 57.63 | 68.4 | 94.89 | 98.58 |

TABLE V: Evaluation of the RAG pipeline on the NQ dataset

| Model/Pipeline | EM | F1 | Top-5 | Top-20 |
|---|---|---|---|---|
| GLaM (Oneshot) [12] | 26.3 | | | |
| GLaM (Zeroshot) [12] | 24.7 | | | |
| PaLM540B (Oneshot) [13] | 29.3 | | | |
| Blended RAG (Zero-shot) | 42.63 | 53.96 | 88.22 | 88.88 |

| Method | BM25 + Match Query | BM25 + Best Field | KNN + Match Query | KNN + Best Field | ELSER + Match Query | ELSER + Best Field |
|---|---|---|---|---|---|---|
| CoQA | 45.3% | 45.3% | 47.56% | 47.56% | 49.94% | 49.94% |

Fig. 9: Top-5 retrieval accuracy

## VI. DISCUSSION

While RAG is a commonly used approach in the industry, we realized during the course of this study that various challenges still exist, like there are no standard datasets on which both R (Retriever) and RAG benchmarks are available. Retriever is often studied as a separate problem in the IR domain, while RAG is studied in the LLM domain. We thus attempted to bring synergy between the two domains with this work. In this section, we share some learning on limitations and appropriate use of this method.

### A. Trade-off between Sparse and Dense Vector Indices

The HotPotQA corpus presents substantial computational challenges with 5M documents, generating a dense vector index to an approximate size of 50GB, a factor that significantly hampers processing efficiency. Dense vector indexing, characterized by its rapid indexing capability, is offset by a relatively sluggish querying performance. Conversely, sparse vector indexing, despite its slower indexing process, offers expeditious querying advantages. Furthermore, a stark contrast in storage requirements is observed; for instance, the sparse vector index of the HotPotQA corpus occupied a mere 10.5GB as opposed to the 50GB required for the dense vector equivalent.

In such cases, we recommend sparse encoder indexes. Furthermore, for enterprises with this volume, we found it better to use multi-tenancy with federated search queries.

### B. Blended Retrievers without Metadata

When datasets are enriched with metadata or other relevant informational facets, they improve the efficacy of blended retrievers. Conversely, for datasets devoid of metadata, such as CoQA, it is not as impressive.

The absence of metadata in the CoQA dataset resulted in hybrid queries offering no improvement over basic queries. This limitation underscores the critical role of metadata in enhancing the efficacy of complex query structures. However, Sparse Encoder-based semantic searches still yield the most favorable outcomes than traditional methods.

Additionally, we would like to note that while NDCG@10 scores for Retriever and F1,EM scores for RAG are commonly used metrics, we found them to be poor proxies of Generative Q&A systems for human alignment. Better metrics to evaluate the RAG system is a key area of future work.

## VII. CONCLUSION

Blended RAG pipeline is highly effective across multiple datasets despite not being specifically trained on them. Notably, this approach does not necessitate exemplars for prompt engineering which are often required in few-shot learning, indicating a robust generalization capability within the zero-shot paradigm. This study demonstrated:

- Optimization of R with Blended Search: Incorporating Semantic Search, specifically Sparse Encoder indices coupled with 'Best Fields' queries, has emerged as the superior construct across all, setting a new benchmark of 87% for Retriever Accuracy on TREC-COVID.
- Enhancement of RAG via Blended Retrievers: The significant amplification in retrieval accuracy is particularly pronounced for the overall evaluation of the RAG pipeline, surpassing prior benchmarks on fine-tuned sets by a wide margin. Blended RAG sets a new benchmark at 68% F1 Score on SQUAD and 42% EM Score on NQ dataset; for non-tuned Q&A systems.

The empirical findings endorse the potency of Blended Retrievers in refining RAG systems beyond focusing on LLM size & type, getting better results with relatively smaller LLM and thus setting a foundation for more intelligent and contextually aware Generative Q&A systems.

ACKNOWLEDGMENT

Authors would like to acknowledge the below members for making this study possible.

- **IBM Ecosystem** The authors conducted this study while employed at IBM Ecosystem. They would like to express their gratitude to the Ecosystem team and leadership for their support in carrying out this work.
- **IBM Research** The authors have received generous feedback on their work from colleagues at IBM Research, particularly Radu Florian, whom the authors would like to acknowledge.
- **Elastic** - The authors have been granted access to the Elastic Search platform and ELSER index as an embodiment of sparse index. They would like to thank Elastic for their support.

# APPENDIX A
## DATASET INFORMATION

### A. TREC-COVID dataset

The TREC-COVID [6]Challenge leveraged the continually updated COVID-19 Open Research Dataset (CORD-19) to assess information retrieval systems, providing valuable insights for the current pandemic response and future system development. This iterative evaluation process, involving biomedical experts' relevance judgments, culminated in a comprehensive test collection known as TREC-COVID Complete, facilitating research on dynamic search environments. Table 6 shows the basic structure of the Trec-Covid dataset.

TABLE VI: Trec-Covid Dataset

| id | title | text | metadata |
|---|---|---|---|
| ug7v899j | Clinical features of culture-proven Mycoplasma | OBJECTIVE: This retrospective chart review des | {'url': 'https://www.ncbi.nlm.nih.gov/pmc/arti' |
| 02tnwd4m | Nitric oxide: a pro-inflammatory mediator in l | Inflammatory diseases of the respiratory tract | {'url': 'https://www.ncbi.nlm.nih.gov/pmc/arti' |
| ejv2xln0 | Surfactant protein-D and pulmonary host defense | Surfactant protein-D (SP-D) participates in th | {'url': 'https://www.ncbi.nlm.nih.gov/pmc/arti' |
| 2b73a28n | Role of endothelin-1 in lung disease | Endothelin-1 (ET-1) is a 21 amino acid peptide | {'url': 'https://www.ncbi.nlm.nih.gov/pmc/arti' |
| 9785vg6d | Gene expression in epithelial cells in respons | Respiratory syncytial virus (RSV) and pneumoni | {'url': 'https://www.ncbi.nlm.nih.gov/pmc/arti' |

### B. NQ dataset

We downloaded this data from GitHub Bier. This dataset is created by Google AI, which is available for open source to help out open-domain question answering; the NQ corpus contains questions from real users, and it requires QA systems to read and comprehend an entire Wikipedia article that may or may not contain the answer to the question. Table 7 shows the basic structure of NQ dataset [5].

TABLE VII: NQ Dataset

| id | text | relevant | answers |
|---|---|---|---|
| 153 | what episode in victorious is give it up | 1 | Freak the Freak Out |
| 7043 | malcolm in the middle what is their last name | 14 | Wilkerson |
| 4392 | distance from las vegas to red wood forest | 16 | 15 miles |
| 8260 | what kind of animal is boots from dora | 18 | anthropomorphic monkey |
| 6740 | where did the rockefeller tree come from 2014 | 21 | Danville , PA |

### C. HotpotQA dataset

In order to enable more explainable question answering systems, HotpotQA [7] is a question answering dataset with natural, multi-hop questions and strong supervision for supporting facts. A group of NLP researchers from Université de Montréal, Stanford University, and Carnegie Mellon University gathers it. Table 8 shows the basic structure of the HotpotQA dataset.

### D. CoQA dataset

The CoQA dataset [10] is an open-source, large-scale dataset for building conversational question-answering systems. The goal of the CoQA challenge is to measure the ability of machines to understand a text passage and answer a series of interconnected questions that appear in a conversation. This data contains dev and train datasets.

Train dataset is used to fine-tune the model. In our experiments, we used a dev dataset. This dataset has 500 documents concerning 7983 question-answer pairs. Table 9 shows the basic structure of the database.

### E. SQuAD Dataset

The Stanford Question Answering Dataset (SQuAD) [9] dataset is an open-source large-scale dataset. It is a collection of question-and-answer pairs derived from Wikipedia articles. There are two datasets available: squad 1.1 and Squad 2.0. We used the squad 1.0 dataset.

This data contains dev and train datasets. We used the dev dataset for our experiments. This dataset contains 2067 documents and 10570 question-answer pairs. Table 10 shows the basic structure of the database.

TABLE VIII: HotpotQA Dataset

| _id | title | text | metadata |
|---|---|---|---|
| 12 | Anarchism | Anarchism is a political philosophy that advocates self-governed societies based on voluntary institutions. These are often described as stateless societies, although several authors have defined them more specifically as institutions based on non-hie | 'url': 'https://en.wikipedia.org/wiki?curid=12' |
| 25 | Autism | Autism is a neurodevelopmental disorder characterized by impaired social interaction, impaired verbal and non-verbal communication, and restricted and repetitive behavior. Parents usually notice signs in the first two years of their child's life. Thes | 'url': 'https://en.wikipedia.org/wiki?curid=25' |
| 39 | Albedo | Albedo is a measure for reflectance or optical brightness (Latin albedo, whiteness) of a surface. It is dimensionless and measured on a scale from zero (corresponding to a black body that absorbs all incident radiation) to one (corresponding t | 'url': 'https://en.wikipedia.org/wiki?curid=39' |
| 290 | A | A (named , plural "As", "A's", "a"s, "a's" or "aes" ) is the first letter and the first vowel of the ISO basic Latin alphabet. It is similar to the Ancient Greek letter alpha, from which it derives. The upper-case version consists of the two slanting | 'url': 'https://en.wikipedia.org/wiki?curid=290' |
| 303 | Alabama | Alabama ( ) is a state in the southeastern region of the United States. It is bordered by Tennessee to the north, Georgia to the east, Florida and the Gulf of Mexico to the south, and Mississippi to the west. Alabama is the 30th largest by area and th | 'url': 'https://en.wikipedia.org/wiki?curid=303' |

TABLE IX: CoQA Dataset

| version | data |
|---|---|
| 1 | {'source': 'wikipedia', 'id': '3zotghdk5ibi9cex97fepx7jetpso7', 'filename': 'Vatican_Library.txt', 'story': 'The Vatican Apostolic Library (), more commonly called the Vatican Library or simply the Vat, is the library of the Holy See, located in Vatican City. Formally established in 1475, although it is much older, it is one of the oldest libraries in the world and contains one of the most significant collections of historical texts. It has 75,000 codices from throughout history, as well as 1.1 million printed books, which include some 8,500 incunabula. The Vatican Library is a research library for history, law, philosophy, science and theology. The Vatican Library is open to anyone who can document their qualifications and research needs. Photocopies for private study of pages from |
| 1 | {'source': 'cnn', 'id': '3wj1oxy92agboo5nlq4r7bndc3t8a8', 'filename': 'cnn_fe05c61a7e48461f7883cdec387567029614f07b.story', 'story': 'New York (CNN) – More than 80 Michael Jackson collectibles – including the late pop star's famous rhinestone-studded glove from a 1983 performance – were auctioned off Saturday, reaping a total $2 million. Profits from the auction at the Hard Rock Cafe in New York's Times Square crushed pre-sale expectations of only $120,000 in sales. The highly prized memorabilia, which included items spanning the many stages of Jackson's career, came from more than 30 fans, associates and family members, who contacted Julien's Auctions to sell their gifts and mementos of the singer. Jackson's flashy glove was the big-ticket item of the night, fetching $420,000 |
| 1 | {'source': 'gutenberg', 'id': '3bdcf01ogxu7zdn9vlrbf2rqzwplyf', 'filename': 'data/gutenberg/txt/Zane Grey___Riders of the Purple Sage.txt/CHAPTER VII_78c077ef5e268383edbec1f1c9d644b1423f889d258d95ff055aa92', 'story': 'CHAPTER VII. THE DAUGHTER OF WITHERSTEEN "Lassiter, will you be my rider?" Jane had asked him. "I reckon so," he had replied. Few as the words were, Jane knew how infinitely much they implied. She wanted him to take charge of her cattle and horse and ranges, and save them if that were possible. Yet, though she could not have spoken aloud all she meant, she was perfectly honest with herself. Whatever the price to be paid, she must keep Lassiter close to her; she must shield from him the man who had led Milly Erne to Cottonwoods. In her fear she so controlled her mind |
| 1 | {'source': 'cnn', 'id': '3ewijtffvo7wwchw6rtyaf7mfwte0p', 'filename': 'cnn_0c518067e0df811501e46b2e1cd1ce511f1645b7.story', 'story': '(CNN) – The longest-running holiday special still has a very shiny nose. "Rudolph the Red-Nosed Reindeer" premiered on television December 6, 1964, and is now one of the holiday season's perennial favorites. The story of the reindeer who saves Christmas is beloved among children and adults alike. The Rankin-Bass animated film production company used Japanese puppets and stop motion to tell the tale, bolstered by a soundtrack featuring Burl Ives' rendition of the theme song. In the story, Santa's reindeer Donner and his wife have a son, Rudolph, who has the distinction of a nose that glows. He runs away after being made to feel an outcast and links.. |

# APPENDIX B
## INDEX AND QUERIES DETAILS

### A. BM25 Index

Fundamental to our exploration is the BM25 index [3], which utilizes TF-IDF (Term Frequency-Inverse Document Frequency) to evaluate document relevance based on query terms. This index serves as a cornerstone for our initial forays into search queries, providing a robust alternative to mere keyword-centric searches. The BM25 index is particularly adept at employing full-text search capabilities enhanced by fuzzy matching techniques, laying the groundwork for more sophisticated query operations.

TABLE X: Squad Dataset

| data | version |
|---|---|
| {'title': 'Super_Bowl_50', 'paragraphs': [{'context': 'Super Bowl 50 was an American football game to determine the champion of the National Football League (NFL) for the 2015 season. The American Football Conference (AFC) champion Denver Broncos defeated the National Football Conference (NFC) champion Carolina Panthers 24–10 to earn their third Super Bowl title. The game was played on February 7, 2016, at Levi's Stadium in the San Francisco Bay Area at Santa Clara, California. As this was the 50th Super Bowl, the league emphasized the "golden anniversary" with various gold-themed initiatives, as well as temporarily suspending the tradition of naming each Super Bowl game with Roman numerals (under which the game would have been known as "Super Bowl L"), so that the logo could prominent | 1.1 |
| {'title': 'Super_Bowl_50', 'paragraphs': [{'context': 'Super Bowl 50 was an American football game to determine the champion of the National Football League (NFL) for the 2015 season. The American Football Conference (AFC) champion Denver Broncos defeated the National Football Conference (NFC) champion Carolina Panthers 24–10 to earn their third Super Bowl title. The game was played on February 7, 2016, at Levi's Stadium in the San Francisco Bay Area at Santa Clara, California. As this was the 50th Super Bowl, the league emphasized the "golden anniversary" with various gold-themed initiatives, as well as temporarily suspending the tradition of naming each Super Bowl game with Roman numerals (under which the game would have been known as "Super Bowl L"), so that the logo could prominent | 1.1 |
| {'title': 'Super_Bowl_50', 'paragraphs': [{'context': 'Super Bowl 50 was an American football game to determine the champion of the National Football League (NFL) for the 2015 season. The American Football Conference (AFC) champion Denver Broncos defeated the National Football Conference (NFC) champion Carolina Panthers 24–10 to earn their third Super Bowl title. The game was played on February 7, 2016, at Levi's Stadium in the San Francisco Bay Area at Santa Clara, California. As this was the 50th Super Bowl, the league emphasized the "golden anniversary" with various gold-themed initiatives, as well as temporarily suspending the tradition of naming each Super Bowl game with Roman numerals (under which the game would have been known as "Super Bowl L"), so that the logo could prominent | 1.1 |
| {'title': 'Super_Bowl_50', 'paragraphs': [{'context': 'Super Bowl 50 was an American football game to determine the champion of the National Football League (NFL) for the 2015 season. The American Football Conference (AFC) champion Denver Broncos defeated the National Football Conference (NFC) champion Carolina Panthers 24–10 to earn their third Super Bowl title. The game was played on February 7, 2016, at Levi's Stadium in the San Francisco Bay Area at Santa Clara, California. As this was the 50th Super Bowl, the league emphasized the "golden anniversary" with various gold-themed initiatives, as well as temporarily suspending the tradition of naming each Super Bowl game with Roman numerals (under which the game would have been known as "Super Bowl L"), so that the logo could prominent | 1.1 |

We used the BM25 index with match queries and combinations of multi_match queries. Table 11 shows the query examples which we used for our experiments.

TABLE XI: BM25 Queires

| Query Type | Query Syntax | Explanation |
|---|---|---|
| Match Query | "query": {"match": {"content":{ "query": ""} } } } | The match query is the standard query for performing a full-text search, including options for fuzzy matching. |
| Cross-field (Blended Query) | {"query":{"multi_match" :{"query":"", "type":"cross_fields", "analyzer":"standard", "fields":[ "title", "text"]}}} | Treats fields with the same analyzer as though they were one big field. Looks for each word in any field. Fields are either title or content. |
| Most Field | {"query": { "multi_match" : { "query":"", "type":"most_fields", "fields":[ "title", "text"] }}} | Finds documents that match any field and combines the _score from each field. Returns top document as result based on scores. |
| Phrase prefix | {"query": { "multi_match" : { "query":"", "type":"phrase_prefix","fields":[ "title", "text"] }}} | Run a match_phrase_prefix query on each field and uses the _score from the best field. |
| Bool prefix | {"query": { "multi_match" : { "query":"", "type":"bool_prefix","fields":[ "title", "text"] }}} | Creates a match_bool_prefix query on each field and combines the _score from each field |
| Best-fields | {"query":{"multi_match" :{"query":"", "type":"best_fields", "analyzer":"standard", "fields":[ "title", "text"]}}} | Finds documents which match any field, but uses the _score from the best field |

## B. KNN Index

Dense Vector (KNN) Index: In pursuit of enhanced retrieval precision, we construct a dense vector index. This approach, empowered by sentence transformers, advances the retrieval process beyond the traditional confines of keyword and similarity searches. Utilizing k-nearest neighbor [4] (KNN) algorithms, this vector search method proffers a marked improvement in accuracy by identifying the proximity of vector representations derived from document and query content.

We used the KNN index with KNN query + match query and combinations of multi_match queries. Table 12 shows the query examples which we used for our experiments.

TABLE XII: KNN Queries

| Query Type | Query Syntax | Score Calculation |
| --- | --- | --- |
| KNN Simple Query | `"knn": {"field": "vector", "query_vector": [], "k":10, "num_candidates":100, "fields": ["text","title"]}` | Dense vector score |
| KNN + Cross-field (Blended Query) | `search_query1 = {"query": {"multi_match" : {"query":"", "type":"cross_fields", "fields":["title", "text"]}}, "knn": {"field":"vector", "query_vector": [54, 10, -2], "k": 10, "num_candidates": 100, "boost": 0.1}, "size": 10}` | Dense vector score + boost + Multi_match query score |
| KNN + Most Field | `search_query1 = {"query": {"multi_match" : {"query":"", "type":"most_fields", "fields":["title", "text"]}}, "knn": {"field":"vector", "query_vector": [54, 10, -2], "k": 10, "num_candidates": 100, "boost": 0.1}, "size": 10}` | Dense vector score + boost + Multi_match query score |
| KNN + Phrase prefix | `search_query1 = {"query": {"multi_match" : {"query":"", "type":"phrase_prefix", "fields":["title", "text"]}}, "knn": {"field":"vector", "query_vector": [54, 10, -2], "k": 10, "num_candidates": 100, "boost": 0.1}, "size": 10}` | Dense vector score + boost + Multi_match query score |
| KNN + Bool prefix | `search_query1 = {"query": {"multi_match" : {"query":"","type":"bool_prefix","fields":["title", "text"]}},"knn": {"field":"vector", "query_vector": [54, 10, -2], "k": 10,"num_candidates": 100,"boost": 0.1},"size": 10}` | Dense vector score + boost + Multi_match query score |
| KNN + Best field | `search_query1 = {"query": { "multi_match" : {"query":"","type":"best_fields", "fields":[ "title", "text"]}},"knn": {"field":"vector", "query_vector": [54, 10, -2], "k": 10,"num_candidates": 100,"boost": 0.1},"size": 10}` | Dense vector score + boost + Multi_match query score |

## C. Sparse Encoder Model Index

Sparse Encoder (Sparse EncodeR Retriever Model) Index: The Sparse EncodeR Retriever Model index is emblematic of our commitment to semantic search. This model, an amalgam of semantic understanding and similarity-based retrieval, allows for the fusion of these paradigms to formulate hybrid queries. By harnessing this index, we elevate our search capabilities to encapsulate the nuanced relationships between terms, thereby capturing a more authentic representation of user intent and document relevance.

We used the Sparse Encoder Model-based index with sparse encoder query + match query and combinations of multi_match queries. Table 13 shows the query examples which we used for our experiments.

TABLE XIII: Query Types and Score Calculations

| Query Type | Query Syntax | Score Calculation |
|---|---|---|
| Sparse EncodeR Retriever Model(SERM) Match Query | `"query":{"text_expansion":{"ml.tokens":{"model_id":".Sparse EncodeR retriever model_model_1","model_text":""}}}` | Sparse vector score |
| Sparse EncodeR Retriever Model + Cross field | `{"query": {"bool": {"should": [{"text_expansion": {"ml.tokens": {"model_text": "", "model_id":".Sparse EncodeR retriever model_model_1"}}}], "must": {"multi_match" : {"query": "", "type": "cross_fields", "analyzer": "standard", "fields": [ "title", "text"]}}}}}` | Sparse vector score + boost + Multi_match query score |
| Sparse EncodeR Retriever Model + Most Field | `"query": {"bool": {"should": [{"text_expansion": {"ml.tokens": {"model_text": "", "model_id":".Sparse EncodeR retriever model_model_1"}}}], "must": {"multi_match" : {"query": "", "type": "most_fields", "fields": [ "title", "text"]}}}}` | Sparse vector score + boost + Multi_match query score |
| Sparse EncodeR Retriever Model + Phrase prefix | `"query": {"bool": {"should": [{"text_expansion": {"ml.tokens": {"model_text": "", "model_id":".Sparse EncodeR retriever model_model_1"}}}], "must": {"multi_match" : {"query": "", "type": "phrase_prefix", "fields": [ "title", "text"]}}}}` | Sparse vector score + boost + Multi_match query score |
| Sparse EncodeR Retriever Model + Bool prefix | `"query": {"bool": {"should": [{"text_expansion": {"ml.tokens": {"model_text": "", "model_id":".Sparse EncodeR retriever model_model_1"}}}], "must": {"multi_match" : {"query": "", "type": "bool_prefix", "fields": [ "title", "text"]}}}}` | Sparse vector score + boost + Multi_match query score |
| Sparse EncodeR Retriever Model + Best field | `"query": {"bool": {"should": [{"text_expansion": {"ml.tokens": {"model_text": "", "model_id":".Sparse EncodeR retriever model_model_1"}}}], "must": {"multi_match" : {"query": "", "type": "best_fields", "fields": [ "title", "text"]}}}}` | Sparse vector score + boost + Multi_match query score |

TABLE XIV: List of Abbreviations

| Abbreviation | Full Form |
|---|---|
| KNN | k-nearest neighbour |
| MQ | Match Query |
| BF | Best Fileds |
| SERM | Sparse EncodeR Retriever Model |
| NQ | Natural Questions |
| EM | Exact Match |

APPENDIX C
ABBREVIATION TABLE

APPENDIX D
RETRIEVER RESULTS- TOP-K ACCURACY RESULTS

The following section encapsulates the retrieval accuracy of our evaluative approach, quantified by Top-k metrics where $k \in \{5, 10, 20\}$, across various datasets:
1) NQ (Natural Questions) dataset
2) TREC-Covid dataset
3) SQuAD (Stanford Question Answering Dataset)
4) CoQA (Conversational Question Answering)

Tables 14, 15, and 16 synopsis is designed to demonstrate the comprehensive capability of our retrieval methodology within diverse informational contexts.

It can be concluded from the results that 'Blended Retriever' offers better accuracy than current methods across all the datasets. Sparse EncodeR Retriever Model (SERM) Based index with Best field queries often given best results with 88% top-5 accuracy for NQ-Dataset and 94% on TREC-Covid. The numbers increase for Top-10 and Top-20 accuracy. Figure 10, Figure 11, and Figure 12 show all these results.

TABLE XV: Top-5 Retrieval accuracy

| Top-5 retrieval accuracy | BM25 + MQ | BM25+ BF | KNN + MQ | KNN + BF | SERM + MQ | SERM + BF |
|---|---|---|---|---|---|---|
| NQ Dataset | 25.19 | 85.05 | 87 | 87.67 | 88 | 88.22 |
| Trec-covid Score1 | 36 | 40 | 36 | 40 | 46 | 48 |
| Trec-covid Score2 | 86 | 86 | 86 | 92 | 92 | 94 |
| HotpotQA | 49.52 | 52.28 | | | | |
| SqUAD | 91.5 | 91.52 | 94.86 | 94.89 | 90.7 | 90.7 |

TABLE XVI: Top-10 Retrieval accuracy

| Top-10 retrieval accuracy | BM25 + MQ | BM25+ BF | KNN + MQ | KNN + BF | SERM + MQ | SERM+ BF |
|---|---|---|---|---|---|---|
| NQ Dataset | 36.7 | 86.26 | 88.46 | 88.66 | 88.55 | 88.77 |
| Trec-covid Score1 | 66 | 72 | 66 | 74 | 52 | 78 |
| Trec-covid Score2 | 92 | 96 | 96 | 97 | 64 | 98 |
| HotpotQA | 55 | 58.93 | | | 62.5 | 65.7 |
| SqUAD | 94.43 | 94.49 | 97.43 | 97.43 | 94.13 | 94.16 |

TABLE XVII: Top-20 Retrieval accuracy

| Top-20 retrieval accuracy | BM25 + MQ | BM25+ BF | KNN + MQ | KNN + BF | SERM + MQ | SERM + BF |
|---|---|---|---|---|---|---|
| NQ Dataset | 37.13 | 87.12 | 88.58 | 88.66 | 88.66 | 88.88 |
| Trec-covid Score1 | 86 | 90 | 90 | 92 | 94 | 98 |
| Trec-covid Score2 | 98 | 100 | 100 | 100 | 100 | 100 |
| HotpotQA | 61.32 | | | | | |
| SqUAD | 96.3 | 96.36 | 98.57 | 98.58 | 96.49 | 96.52 |

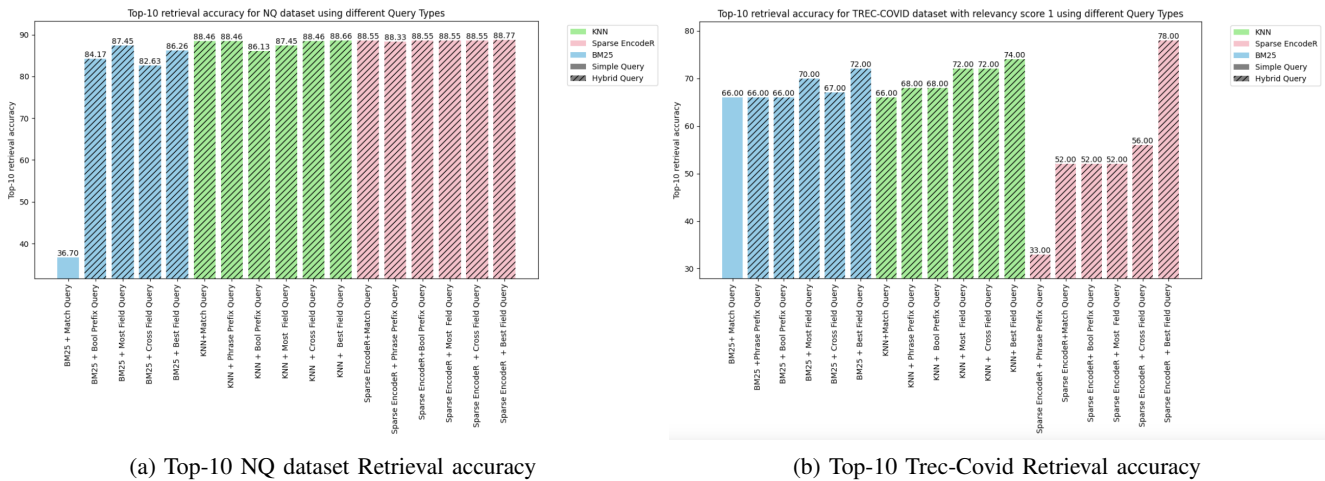

(a) Top-10 NQ dataset Retrieval accuracy  (b) Top-10 Trec-Covid Retrieval accuracy

Fig. 10: Top-10 Retrieval accuracy

APPENDIX E
RAG EVALUATION RESULTS

Distinctively, our Blended RAG approach has not undergone training on any related corpora. It harnesses an optimized amalgamation of field selections, query formulations, indices, and Large Language Models (LLMs) to render the most precise responses possible. We used FlanT5-XXL for this pipeline. Consequently, the Blended RAG showcases enhanced performance in the RAG use case, even without dataset-specific fine-tuning. This characteristic renders it particularly advantageous for large enterprise datasets, where fine-tuning may be impractical or unfeasible, underscoring this research's principal application.

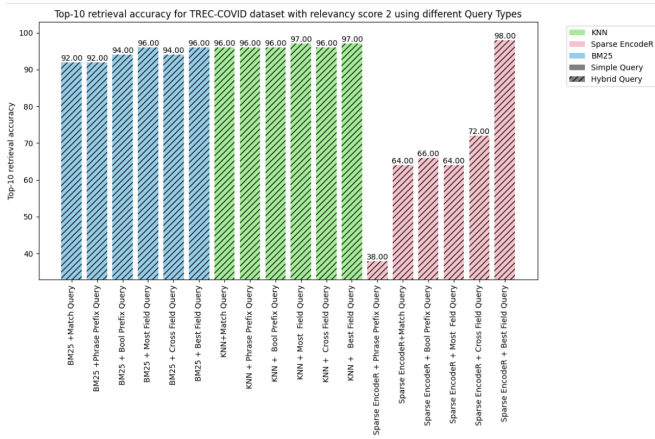
(a) Top-10 Trec Covid Score 2 Retrieval accuracy

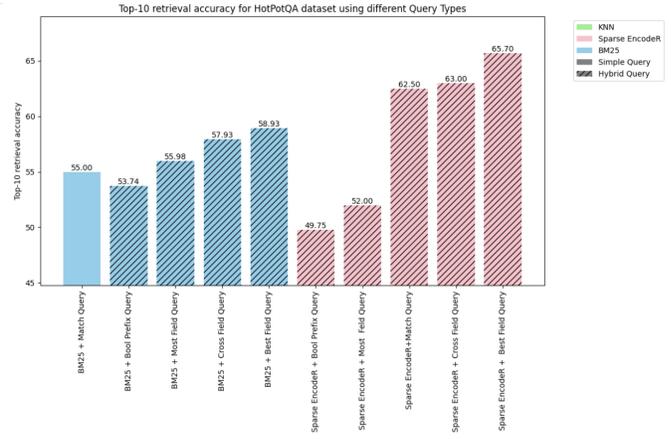
(b) Top-10 HotpotQA Retrieval accuracy

Fig. 11: Top-10 Retrieval accuracy

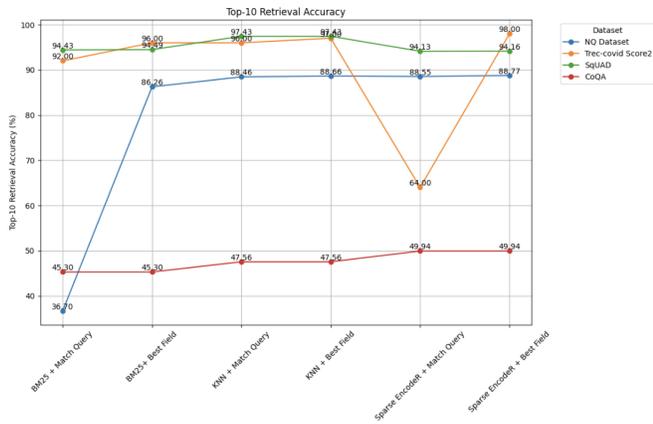
(a) Top-10 Retrieval accuracy

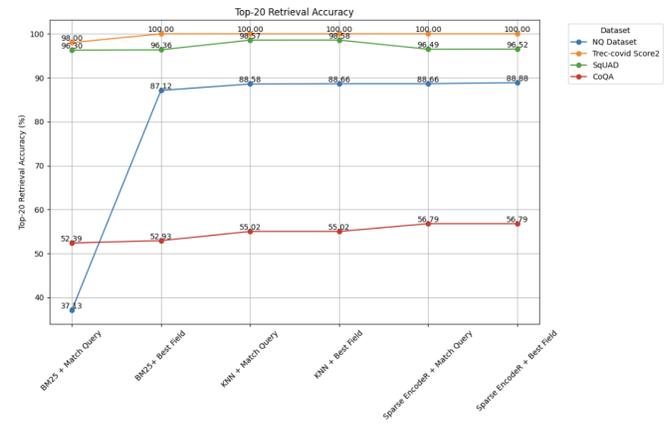
(b) Top-20 Retrieval accuracy

Fig. 12: Top-k Retrieval accuracy

Table 17 shows the RAG evaluation results for the NQ Dataset with all relevant matrices.

We added more metrics for RAG evaluations than just the F1 scores to include additional scores like BLUE, METEOR, ROUGUE, SIM-HASH, PERPLEXITY, BLUERT, BERT. We found that useful metrics differ by use case and dataset nature.

TABLE XVIII: RAG Evaluation on NQ Dataset

| Query Types | EM | F1 | blue_score | meteor_score | rouge_score | sentence_similarity | sim_hash | perplexity_score | bleurt_score1 | bert_score |
|---|---|---|---|---|---|---|---|---|---|---|
| BM25 + MQ | 32.91 | 40.4 | 3.81 | 33.47 | 42.65 | 57.47 | 18.95 | 3.15 | 27.73 | 6.11 |
| BM25+ BF | 37.58 | 47.31 | 4.63 | 3.98 | 49.79 | 63.33 | 17.02 | 3.07 | 13.62 | 65.11 |
| KNN + MQ | 40.21 | 50.51 | 4.77 | 42.11 | 53.32 | 67.02 | 15.94 | 3.04 | 5.12 | 67.27 |
| KNN + BF | 40.32 | 50.45 | 5.05 | 42.34 | 53.24 | 66.88 | 15.94 | 3.048 | 5.7 | 67.3 |
| ELSER + MQ | 42.63 | 53.96 | 5.27 | 45.13 | 57.07 | 70.47 | 14.95 | 3.01 | 2.02 | 69.25 |
| ELSER + BF | 42.3 | 53.25 | 5.24 | 44.77 | 56.36 | 69.65 | 15.14 | 3.02 | 0.24 | 68.97 |

## APPENDIX F
## GITHUB REPO

The GitHub Repo for this work is https://github.com/ibm-ecosystem-engineering/Blended-RAG